\documentstyle[12pt]{article}
\begin{document}
\title{THE SYMMETRY UNDERLYING SPIN AND THE DIRAC EQUATION$^+$}
\author{B.G. Sidharth$^*$\\
Centre for Applicable Mathematics \& Computer Sciences\\
B.M. Birla Science Centre, Adarsh Nagar, Hyderabad - 500063 (India)}
\date{}
\maketitle
\footnotetext{\noindent E-mail:birlasc@hd1.vsnl.net.in\\
$^+$Paper at the Spin and Perturbation Theory 98, symposium, Rome, 1998.}
\begin{abstract}
It is shown, in the context of a recent formulation of elementary particles
in terms of, what may be called, a Quantum Mechanical Kerr-Newman metric,
that spin is a consequence of a space-time cut off at the Compton wavelength
and Compton time scale. On this basis, we deduce the Dirac equation from a
simple coordinate transformation.
\end{abstract}
\section{Introduction}
The spinorial behaviour of elementary particles is a great divide which
stands in the way of the unification of General Relativity and Quantum
Mechanics. As noted by Wheeler and others\cite{r1,r2}, the problem of
introducing spin half into General Relativity, or in the reverse, introducing
the curvature of space into the description of the electron had defied solution
for several decades. While the concept of spin half was introduced by
Uhlenbeck and Goudsmit,
it is the Dirac equation which brings out the Quantum Mechanical equation of
the electron in conformity with Special Relativity\cite{r3}, wherein we see
the emergence not only of the spin, but also the anomalous gyro magnetic
ratio. As is well known the starting point for the Dirac equation is the
square root operator which translates itself into the well known four
dimensional representation indicating the inner or extra degrees of
freedom:
$$H = \sqrt (c^2p^2 + m^2c^4),$$
where
$$H = \imath \hbar \frac{\partial}{\partial t} \mbox{and} \vec p =
-\imath \hbar \vec \nabla$$
However, while the Dirac equation is relativistically covariant, the square
root operator masks any underlying fundamental symmetry.\\
It is ofcourse true that one could invoke the spinorial representation of the
rotation group, but the question is, why the $\underline{spinorial}$ representation?
Our usual three dimensional space would suggest that the ordinary representation
connected with the orbital angular momentum would be the natural choice. But
this is not so.\\
In this context we will now show that if we invoke quantized space time, we are
lead to the Dirac equation by a simple coordinate transformation in Minkowski space.
\section{Space Time Quantization}
It was suggested in the context of a Quantum Mechanical Kerr-Newman formulation
of electrons that space time has in the final analysis a discrete character\cite{r4,r5},
and that indeed this could be more fundamental than quantized energy\cite{r6,r7}.
Infact a simple way to see this is that in the theory of the Dirac equation,
the position operator is non Hermitian, that is the position coordinates
are complex (cf. ref.\cite{r3}):\\
$$\vec x = c^2\vec p H^{-1} + \frac{\imath}{2} c \hbar (\vec \alpha - c \vec p
H^{-1})H^{-1}$$
The imaginary or rapidly oscillating or
Zitterbewegung part is eliminated, that is the position operator becomes Hermitian,
only on averaging over space time intervals of the order of the Compton
wavelength and Compton time. As we approach the Compton wavelength or time,
we encounter unphysical effects, like negative energy solutions and the
related Zitterbewegung. Indeed this is also true in the classical theory\cite{r8},
and it was shown that by introducing the idea of a Chronon, the minimum unit
of time equalling the Compton time the difficulty could be
circumvented\cite{r9}. Subsequently the concept of the chronon was investigated
by others\cite{r10}.\\
In any case such a space time cut off leads to a not only pleasing description
of spin half particles (cf.ref.\cite{r4,r5}), but also to a consistent
cosmology\cite{r11} which predicts an ever expanding universe as indeed
has recently been confirmed by independent observations.\\
In a more general context, it was shown a long time ago by Snyder\cite{r12,r13}
that discrete space time is entirely compatible with the Laurentz Transformation,
and leads to a divergence free electrodynamics. Given a natural unit length,
$a$, he showed that
$$[x, y] = (\imath a^2 / \hbar)L_{z,}  [t, x] = (\imath a^2 / \hbar c)M_{x,}$$
\begin{equation}
[y, z] = (\imath a^2 / \hbar) L_{x,} [t, y] = (\imath a^2 / \hbar c)M_{y,}\label{e1}
\end{equation}
$$[z, x] = (\imath a^2 / \hbar) L_{y,} [t, z] = (\imath a^2 / \hbar c)M_{z,}$$
where $L_x$ etc. have their usual significance.\\
Similarly it also follows that
$$[x, p_x] = \imath \hbar [1+(a/\hbar)^2 p^2_x];$$
$$[t, p_t] = \imath \hbar [1-(a/ \hbar c)^2 p^2_t];$$
\begin{equation}
[x, p_y] = [y, p_x] = \imath \hbar (a/ \hbar)^2 p_xp_y ;\label{e2}
\end{equation}
$$[x, p_t] = c^2[p_{x,} t] = \imath \hbar (a/ \hbar)^2 p_xp_t ;\mbox{etc}.$$
where $p_x$ etc. have the usual meaning.\\
The surprising thing about equation (\ref{e1}) is that the coordinates do not
commute: We are automatically lead to a non commutative geometry. Equation
(\ref{e2}) also shows a modification to the usual commutation relations. However
as $a \to 0$ we recover the usual theory.
\section{The Emergence of Spin}
In the context of the Quantum Mechanical Kerr-Newman Black Hole formulation
discussed in references\cite{r2,r4,r5} let us consider the special case where
$a = \hbar/mc$, the Compton wavelength and $p_x = mc$ and so on. Substitution
in equation (\ref{e2}) leads to
\begin{equation}
[x, p_x] = 2 \imath \hbar\label{e3}
\end{equation}
and similar equations.\\
The right hand side of (\ref{e3}) is double the usual value - as if the spin has been halved or
the coordinate doubled. This is the typical double connectivity of spin half.\\
This is suggestive of the surprising fact that Quantum Mechanical spin could
arise from quantized space time. This can be confirmed as follows: Let us
consider a transformation of the wave function by a linear operator
$U(R)$, that is
\begin{equation}
|\psi' > = U(R) |\psi >\label{e4}
\end{equation}
If the transformation $R$ is a simple infinitesimal  coordinate shift in Minkowski space
we will get, from (\ref{e4}),  as is well known\cite{r14,r15}
\begin{equation}
\psi' (x_j) = [1 + \imath \epsilon (\imath \epsilon_{ljk} x_k \frac{\partial}
{\partial x_j}) + 0 (\epsilon^2)] \psi (x_j)\label{e5}
\end{equation}
We next consider the commutation relations (\ref{e1}), taking $a$ to be the
Compton wavelength. One can easily verify that the choice
\begin{equation}
t =  \left(\begin{array}{l}
        1 \quad 0 \\ 0 \quad -1 \\
        \end{array}\right), \vec x =
   \ \  \left(\begin{array}{l}
         0 \quad \vec \sigma \\ \vec \sigma \quad 0 \\
       \end{array}\right)\label{e6}
\end{equation}
provides a representation for the coordinates
$x$ and $t$. Equation (\ref{e6}) is precisely a representation
of the Dirac $\gamma$ matrices. Substitution of (\ref{e6}) in (\ref{e5}) now immediately
leads to the Dirac equation.\\
Thus we have shown that given minimum space time intervals of the order of
the Compton wavelength and time, a simple coordinate transformation leads to the Dirac
equation. (Alternatively, we could have considered the operator
$\exp (-\imath\ \epsilon^\mu p_\mu)$.)\\
Ofcourse the Dirac equation itself is meaningful after an averaging over
these minimum space time intervals, thus completing the circle.\\
\section{Discussion}
The above conclusion should not come as a surprise. As argued in
references\cite{r2,r4,r5}, it is the Zitterbewegung effects within the
minimum space time intervals which are symptomatic of the double connectivity
of space or vice versa. Space time {\it points} are classical concepts, which have
no Quantum Mechanical validity, as has been pointed out by several
scholars\cite{r16} - the Heisenberg Uncertainity relation forbids a physical
interpretation of space time points.\\
On the other hand it is precisely this double connectivity of spin half, which
as pointed out has forbidden a reconciliation with the purely classical General
Relativity. Infact a non commutative geometry as a possible solution has
also been speculated upon\cite{r17}.\\
It has also been pointed out that such a space
time cut off would be a solution to the divergences encountered in Quantum
Field Theory (cf.ref.\cite{r4,r13}). Indeed, we are automatically lead to a
quantized electrodynamics.\\
Finally it may be pointed out that in the Kerr-Newman formulation referred
to earlier, inertial mass, electromagnetism and gravitation naturally arise.
In particular, it was argued that electromagnetism is the direct result of the
opposite parity of the upper and lower (or positive and negative) components
of the Dirac four spinor.\\
All this can be seen to be a consequence of space time quantization.

\end{document}